\documentclass[fleqn,a4]{article}
\usepackage{amsmath,amssymb,amscd}
\usepackage[pdftex]{graphicx}
\makeatletter

\selectfont

\renewcommand{\section}{\@startsection{section}{1}{\z@}%
{2ex}{1ex}{\reset@font\large\bfseries}}%
\renewcommand{\thesection}{\@arabic\c@section}
\def\@listi{\topsep=.3\baselineskip \parsep=.2ex \partopsep=0ex%
\itemsep=0ex \leftmargin=4ex \rightmargin=2ex}
\let\@listI\@listi
\@listi\def\@listii{\parsep=.2ex \partopsep=0pt \itemsep=0ex%
\leftmargin=4ex \rightmargin=0ex}
\let\@listiii\@listii
\let\@listiv\@listii
\let\@listv\@listii
\let\@listvi\@listii
\long\def\@makecaption#1#2{\footnotesize\sbox\@tempboxa{#1. #2}
\ifdim\wd\@tempboxa >\hsize #1. #2\par
\else \global\@minipagefalse
\hb@xt@\hsize{\hfil\box\@tempboxa\hfil}
\fi}
\makeatother
\usepackage{graphicx}
\usepackage{amsmath}
\usepackage{url}

\usepackage{textcomp}
\usepackage{xcolor}
\usepackage{color}
\usepackage[geometry]{ifsym}

\begin{document}
\title{Educational Effects in Mathematics: Conditional Average Treatment Effect depending on the Number of Treatments}

\author{\thanks{tnagai@iuhw.ac.jp} Tomoko Nagai, Takayuki Okuda$^1$, Tomoya Nakamura$^1$, Yuichiro Sato$^2$, Yusuke Sato$^3$, \\
Kensaku Kinjo$^1$, Kengo Kawamura$^4$, Shin Kikuta$^1$, Naoto Kumano-go$^1$\\
International University of Health and Welfare, Japan\\
${}^1$Kogakuin University, Japan\\
${}^2$Waseda University, Japan\\
${}^3$Osaka Institute of Technology, Japan\\
${}^4$Osaka Sangyo University, Japan
} 

\date{}


\maketitle
\begin{abstract}
This study examines the educational effect of the Academic Support Center at Kogakuin University. Following the initial assessment, it was suggested that group bias had led to an underestimation of the Center's true impact. 
To address this issue, the authors applied the theory of causal inference. 
By using T-learner, the conditional average treatment effect (CATE) of the Center's face-to-face (F2F) personal assistance program was evaluated.
Extending T-learner, the authors produced a new CATE function
that depends on the number of treatments (F2F sessions) 
and used the estimated function to predict the CATE performance of F2F assistance. 
\end{abstract}

\section{Introduction}
\label{sec:into}
The Academic Support Center at Japan's Kogakuin University (hereafter, the Center) 
offers {\em face-to-face} (F2F) personal assistance, as well as remote personal assistance, along with basic classes and a variety of on-demand educational materials available on its Learning Management System. 
These support options are open to all students. 
The university operates on a four-quarter system (1Q to 4Q) and conducts a proficiency test for all students at the time of admission in addition to regular examinations for credit-certified subjects 
at the end of each quarter. 
(Appendix \ref{sec:data}) 

We studied the educational effect of the Center, and found that 
the average deviation in the 1Q Differentiation (hereafter, Diff.) regular examinations for F2F assistance users was 46.8, or 3.5 points lower than the 50.3 average of the nonusers group. 
At first glance, it appeared that F2F assistance had no positive educational effect on a student's regular examination performance. 
However, the fact that the average proficiency test deviation for F2F users was lower than that for nonusers suggests the possibility of underestimation due to bias. 
To address this issue, we applied the framework of causal inference, setting the use of F2F as a treatment (Appendix \ref{sec:reason}).

To estimate a Conditional Average Treatment Effect  (CATE) \cite{rubin}, 
meta-learners have been introduced in a binary treatment setting \cite{kunzel}. 
Extending the binary setting,  multiple and multi-level treatments models have been proposed \cite{harada, hu, lech, lin,yutas1}.
 Meta-learners for multi-valued treatments makes it possible to identify the impact of the number of possible treatments on CATE performance \cite{acha}. 
A more extensive discussion of related studies is presented in the Appendix \ref{sec:relate}.
 
The main contributions of the present study can be summarized as follows \cite{tnagai3}. 
(i) Using T-learner, we estimate the CATE of F2F assistance on the 1Q Diff. regular examination. 
(ii) Extending T-learner, we propose a new CATE estimator that depends on the number of treatments, which is learned with variables that include the number of treatments. 
(iii) We predict CATE performance depending on the number of F2F sessions.

\section{Causal Inference Setup }
\label{sec:ci}
Because our decision tree \cite{Breiman} results showed that the F2F branches were the most common, we chose F2F as our main focus (Appendix \ref{sec:tree}). 
In this study, we adopted the potential outcome framework \cite{ney, rubin2} and T-learner \cite{kunzel}.
The treatment corresponds to the use of F2F in 1Q.  
The potential outcome corresponds to the deviation value of 1Q Diff. regular examination.
Regarding input variable ${\bf X}$ for the estimators, we considered two types.  
$X_1$ represents the proficiency test deviation value at the time of admission, which we deal with as a covariate.  $X_2$ represents the number of F2F sessions attended by the student in 1Q, which, for us, corresponds to the number of treatments. 
The number of treatments was included as an input variable 
(Appendix \ref{sec:T}).

The data from $N=$ 1,389 students who took 1Q Diff. examination were analyzed. 
Of this group, 91 students used F2F assistance, while the remaining 1,298 students did not. 
To estimate CATE, we used random forests \cite{forest}, implemented by Python scikit-learn.   


\begin{figure}[t]
  \centering
  \includegraphics[scale=0.95]{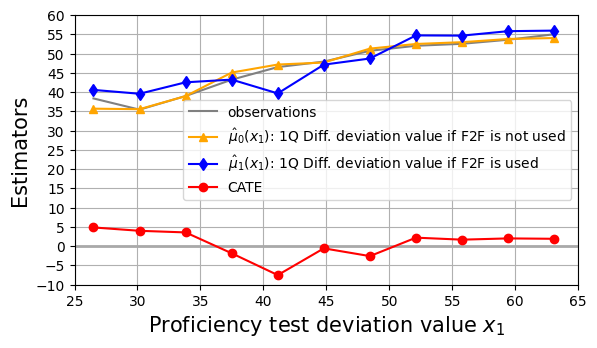}
  \caption{Estimated deviation values vs.\ proficiency test deviation value $X_1=x_1$.}
  \label{fig:CATE}
\end{figure}

\section{Results (Input Variable ${\bf X}, {\bf x} \in \mathbb{ R}^1$)}
\label{sec:resultxr1}
Initially, the input variable ${\bf X}$ was the one-dimensional vector ${\bf X}=(X_1)$. 
We estimated the CATE of F2F as a function of $X_1=x_1$ (Appendix \ref{sec:xr1}). 
Fig. \ref{fig:CATE} shows the estimated values of deviations of 1Q Diff. with $X_1=x_1$. 
The value of $\hat{\mu}_0(x_1)$ is the estimated deviation value of 1Q Diff. if  F2F is not used ($\textcolor{orange}{\blacktriangle}$ in Fig. \ref{fig:CATE}), and $\hat{\mu}_1(x_1)$ is the estimated value if F2F is used 
($\textcolor{blue}{\blacklozenge}$). 
The CATE estimator 
$\hat{\tau} (x_1)$ ($\textcolor{red}{\bullet}$) is defined by 
the difference $( \hat{\mu}_1(x_1)-\hat{\mu}_0(x_1))$.
For $x_1 \le 35 $, 
$\hat{\tau} (x_1)$ ranges from 4 to 5. 
This indicates that using F2F assistance improves the regular examination deviation value in the range of 4 to 5 points. 
$\hat{\tau} (x_1)$ becomes negative for $37 \le x_1 \le 48$ 
but becomes positive again for $x_1\ge 50$, 
showing a value of approximately 2.

Average Treatment Effect ($ATE$) was found to be 0.48 (Appendix \ref{sec:xr1}).
This means that using F2F assistance results in an average improvement of 0.48 in the 1Q Diff. deviation values. 
Thus, the effect of F2F was found to be a 0.48 deviation increase 
rather than the 3.5 deviation decrease described in Section \ref{sec:into}.
Average Treatment Effect on the Treated ($ATT$) was 1.00, 
which is compared to the results in Section \ref{sec:resultxr2}.

\section{Estimation and Results $({\bf X}, {\bf x} \in \mathbb{R}^2)$}
\label{sec:resultxr2}
Multiple regression analysis was used to estimate the potential effect of the number of F2F sessions on test performance (Appendix \ref{sec:dis}). Accordingly, we explicitly added the number of F2F sessions $X_2$ as an input variable. 
The input variable ${\bf X}$ is thus the 2-dimensional vector ${\bf X}=(X_1,X_2)$
(Appendix \ref{sec:xr2}).

Extending the definition of CATE in T-learner, we propose a new CATE estimator 
$\varphi (x_1,x_2 )$ with proficiency test deviation value $X_1=x_1$ and the number of F2F sessions $X_2=x_2$, defined as follows:
\begin{eqnarray}
\varphi (x_1,x_2 ) & = & \frac{1}{|S_{x_1} |}\sum _{k \in S_{x_1}}\{ \hat{\mu}_1 (X_1^{k,obs}, x_2 )  - \hat{\mu}_0 (X_1^{k,obs}, X_2^{k,obs}) \}  \quad (x_2=1,2,\cdots ). 
\label{equ:phi}
\end{eqnarray}
Here, $k$ indicates the $k$-th out of $N$ students, and $S_{x_1}$ is the set of students whose $X_1=x_1$. 
$|S_{x_1}|$  is the number of students in $S_{x_1}$. 
 $X_1^{k, obs}$ and $X_2^{k, obs}$ are the $k$-th student observations of  $X_1^{k}$ and $X_2^{k}$.

\begin{figure}[t]
  \centering
 \includegraphics[scale=0.95]{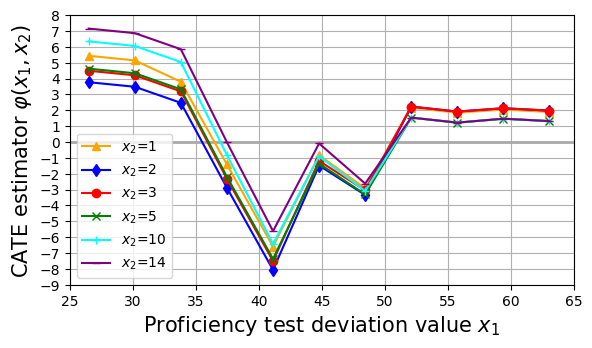}
  \caption{CATE estimator $\varphi (x_1,x_2 )$ vs.\ $x_2$, the number of F2F.}
\label{fig:F2Fd} 
\end{figure}

Fig. \ref{fig:F2Fd} shows the CATE estimator $\varphi (x_1,x_2 )$ in Eq. \eqref{equ:phi} with the number of F2F sessions 
(see also 3D surface plot in Appendix \ref{sec:3D}). 
For proficiency test deviation value $x_1 \leq $ 35, 
$\varphi (x_1, x_2 )$ decreases from $x_2$=1 time ($\textcolor{orange}{\blacktriangle}$ in Fig. \ref{fig:F2Fd}) to $x_2$=2 times ($\textcolor{blue}{\blacklozenge}$). 
However, $\varphi (x_1, x_2 )$ increases as $x_2$ increases from $x_2$=3 times ($\textcolor{red}{\bullet}$), 
5 times ($\textcolor{green}{\times}$), 10 times 
($\textcolor{cyan}{+}$), and 14 times ($\textcolor{purple}{-}$). 
That is to say, we have 
$\varphi (x_1,2) < \varphi (x_1,3) < \varphi (x_1,5) \cdots$. 
It is predicted that three or more F2F sessions contribute to a greater increase  in the deviation value of 1Q Diff. as the number of F2F sessions increases. 
While for $x_1 \geq 50$, $\varphi (x_1,x_2 )$ is roughly 2 if $x_2 \leq 3$ times ($\textcolor{red}{\bullet}$), 
$\varphi (x_1, x_2)$ becomes approximately 1.4 after 5 sessions ($\textcolor{green}{\times}$). 
Although a single F2F session leads to a positive $\varphi(x_1, 1)$, 
it is not predicted that the deviation value of 1Q Diff. increases as the number of F2F sessions increases.

Let $ATT_2$ be the average treatment effect for users with two variables (Appendix \ref{sec:xr2}). 
From the relation of $\hat{\mu}_0 (X_1^{k,obs},X_2^{k,obs})=\hat{\mu}_0 (X_1^{k,obs},0)$, 
$ATT_2$ should be equal to $ATT$. 
In fact, we obtained $ATT_2=1.00$, which is equal to $ATT$=1.00. 

\section{Conclusion}
\label{con}
In this study, we investigated the CATE of the F2F personal assistance offered by Kogakuin University's Academic Support Center on the deviation value of the 1Q Diff. regular examination by using T-learner. 
We proposed a CATE estimator that depends on the number of F2F sessions engaged in by students. We  optimized the estimator with two variables, including the number of F2F sessions. 
Using the CATE estimator, we predicted CATE performance.

\section*{Acknowledgment}
This research was supported by the Japan Society for the Promotion of Science KAKENHI Grant Number JP24K06289.

\appendix

\section{Basic Data}
\label{sec:data}
Below is a summary of the FY2022 data for the Center's Mathematics Division. The data include personal assistance, basic classes, and the Learning Management System of the Mathematics Division at the Center (LMS).

\begin{itemize}
  \item Personal assistance:
Personal assistance sessions were conducted for 40 minutes each. 
From April 6, 2022, to March 27, 2023, a total of 1,278 personal assistance sessions were conducted. These sessions included 1,050 F2F sessions and 228 remote sessions. 
  \item Basic classes:
A number of basic classes were conducted in person during FY2022. A total of 345 students participated in these basic classes. 
The basic class 1Q Diff., corresponding to the 1Q Diff. credit-certified subject, had the highest number of participants (125).
  \item LMS:
On-demand educational materials developed by the Mathematics Division were available on the LMS. 
The materials contained test-type {\em Exercises}, instructional {\em Videos}, and reference materials. 
From April 1, 2022, to March 30, 2023, there were 26,262 total accesses to the LMS. 
\end{itemize}

\section{Reason for using Causal Inference}
\label{sec:reason}

Because Diff. is a mandatory course and has the largest number of participants among all the mathematical subjects, our focus was on the outcomes for this course. 
To ensure uniformity in our analysis, we used deviation values, which were defined as $10\times \frac{  ({\rm score - average })}{{\rm standard~deviation}} + 50$. 

The average deviation for the 91 students who used F2F assistance at least once was 46.8; 
the average for the 1,298 students who did not use F2F assistance was 50.3. 
The average deviation for users was thus 3.5 points lower than that for nonusers. 
At first glance, then, it would appear that F2F has no positive educational effect.
However, the average proficiency test deviation for the users group at the time of admission was 42.7, as compared to
51.4 for nonusers, 
meaning that the average proficiency test deviation of users was 8.7 lower than that of nonusers. 
To us, this suggested a potential underestimation of the effect of F2F assistance due to bias between the two groups. 

\section{Related work}
\label{sec:relate}
Causal Inference is applied in many fields, including medicine \cite{Ala}, economics \cite{knu2}, public policy \cite{ba}, industrial recommender systems \cite{luo, yutas3}, 
and educational sociology \cite{fujihara, kell}. 
One of the main challenges of potential outcome theory \cite{rubin} is to estimate average treatment effects and CATE.

With the development of machine learning, 
machine learning-based models have been proposed for CATE estimation \cite{caronb}. 
K\"unzel introduced the theory of meta-learners (S-learner, T-learner, X-learner) for CATE estimation in a binary treatment setting \cite{kunzel}. 

Recently, the multiple and continuous treatments of Causal Inference have been considered. 
Theoretical work \cite{imbens} extended the propensity score to multi-valued treatment settings.
The causal effects of multiple treatments \cite{lech} and 
multiple treatment uplift modeling \cite{yutas1} have also been discussed, 
and machine-learning based Causal Inference for multi-level treatments has been proposed \cite{lin}. 
Bayesian additive regression trees have been used for estimating the causal effects of multiple treatments\cite{hu}.
In addition, neural networks have been applied to the estimation of individual effects for graph-structured treatments \cite{harada}.

Concerning meta-learners, 
Knaus studied multiple treatment effects in economics\cite{knu}. 
Acharki et al. considered meta-learners for multi-valued treatment heterogeneous effects, which allows an analysis of the impact of the number of possible treatments \cite{acha}. 

In present paper, 
we propose a new CATE function depending on the number of treatments, or more specifically, a CATE estimator that
depends on the number of F2F sessions \cite{tnagai3}. 
The estimator predicts the CATE performance of F2F assistance depending on the number of F2F sessions. 

\section{Decision Tree}
\label{sec:tree}

\begin{figure*}[t]
\begin{center}\hspace{-6cm}\vspace{0cm}
  \includegraphics{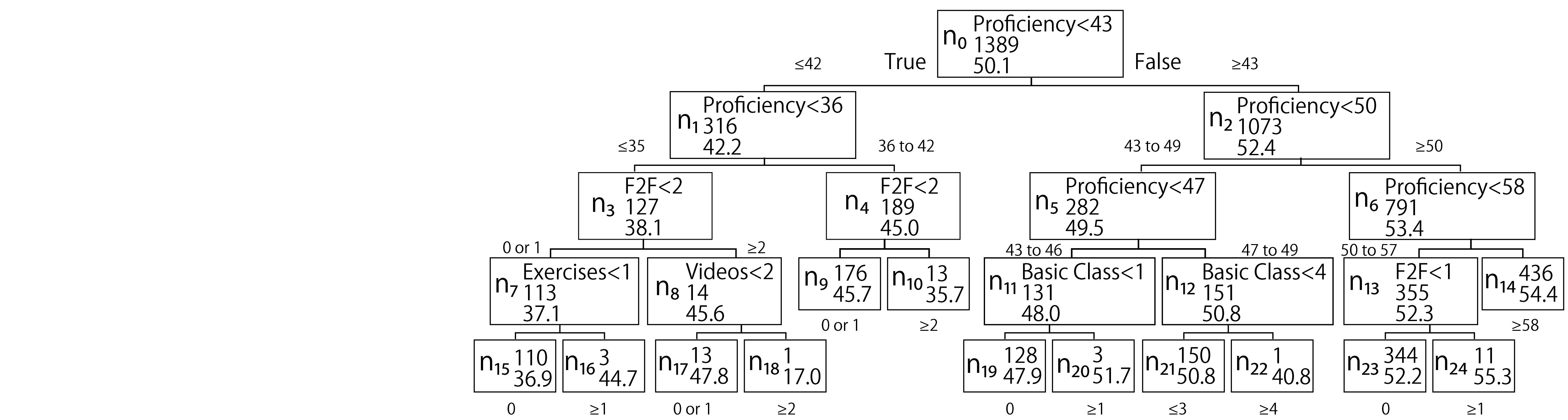}
  \caption{Decision tree for 1Q Differentiation regular examination. 
Each node lists, from top to bottom, the splitting criterion, the number of students, and the average deviation.
The terminal nodes do not have a splitting criterion.}
  \label{fig:tree}
\end{center} 
\end{figure*}

As noted, the Mathematics Division at the Center provides learning resources, including F2F assistance, basic classes, etc. 
To investigate which teaching method influences the deviation values of the 1Q Diff. regular examination 
and to what extent the method influences the examination, we used a decision tree (a regression tree), as was done in previous studies \cite{tnagai1, tnagai2}. 
In this type of decision tree \cite{Breiman}, the input variables that cause higher-level splits indicate stronger influences. 
Our analysis was implemented using Python's scikit-learn.


Fig. \ref{fig:tree} shows a regression tree for the 1Q Diff. regular examination. 
The output variable is the deviation score of the 1Q Diff. regular examination for freshman students. 
The input variables were as follows: \\
\textcircled{\scriptsize1} {\em Proficiency} test deviation value at the time of admission. \\
\textcircled{\scriptsize2} Number of {\em F2F} sessions in 1Q. \\
\textcircled{\scriptsize3} Number of remote personal assistance sessions in 1Q. \\
\textcircled{\scriptsize4} Number of participations in  {\em Basic Class} for 
 Diff. in 1Q.\\ 
\textcircled{\scriptsize5} Number of submissions of {\em Exercises} for Diff. in 1Q. \\
\textcircled{\scriptsize6} Number of connections to {\em Videos} for Diff. in 1Q.\\
\textcircled{\scriptsize7} Number of connections to reference materials for Diff. in 1Q. \\
We removed missing data due to tests not taken. 
Node $n_0$ represents a collection of data from $N$=1,389 students who took the 1Q Diff. regular examination. 
Node $n_0$ is split into $n_1$ (for values $\leq$ 42) and $n_2$ (for values $\geq$ 43) based on the splitting criterion 
{\em Proficiency} $<43$. 
Below are the results related to the Mathematics Division at the Center.

\begin{itemize}
\item 
Node $n_3$ is split based on {\em F2F} $<2$ in the range of {\em Proficiency} $\le$ 35. 
Students who used two or more F2F sessions are directed to the right node $n_8$, 
  while those who used fewer than two are directed to the left node $n_7$. 
The average deviation value of 1Q Diff. regular examination for $n_8$ is 45.6, which is higher than that for $n_7$ of 37.1.
Similarly, node $n_{13}$ is also split based on {\em F2F} $<1$ in the range of {\em Proficiency} 50 to 57. 
The average deviation value of 1Q Diff. for F2F users $n_{24}$ is 55.3, 
 which is higher than that for nonusers $n_{23}$ of 52.2.
\item Node $n_7$ is split based on {\em Exercises} $<1$ 
in the range of {\em Proficiency} $\le$ 35. 
The average deviation value of the 1Q Diff. for submitters $n_{16}$ is 44.7, 
 which is higher than that for nonsubmitters $n_{15}$ of 36.9.
\item 
Node $n_{11}$ is split based on {\em Basic Class} $<1$
in the range of {\em Proficiency} 43 to 46.  
The average deviation value of 1Q Diff. for participants' $n_{20}$ is 51.7, 
 which is higher than that for nonparticipants' $n_{19}$ of 47.9.
 
 \end{itemize}


\section{T-learner}
\label{sec:T}
In this study, treatment corresponds to the use of F2F assistance. 
Potential outcome corresponds to the deviation value of the 1Q Diff. regular examination. 

With respect to the input variable ${\bf X}$ for the estimator,    
$X_1$ represents the proficiency test deviation value at the time of admission, 
and $X_2$ represents the number of F2F sessions used in 1Q. 
In Appendix \ref{sec:xr1} and Section \ref{sec:resultxr1}, 
${\bf X}$ is a one-dimensional vector ${\bf X}=(X_1)$; 
in Appendix \ref{sec:xr2} and Section \ref{sec:resultxr2}, 
${\bf X}$ is a two-dimensional vector ${\bf X}=(X_1, X_2)$. 
$X_1$ corresponds to the covariate variable, and $X_2$ corresponds to the number of treatments. 

Let $Y^k (1)$ be the deviation value of the 1Q Diff. regular examination if student $k$ uses F2F. 
Let $Y^k (0)$ be the deviation of 1Q Diff. if student $k$ does not use F2F assistance. 

The individual treatment effect $D^k$ for student $k$ in the potential outcome model is defined by $D^k=Y^k (1)-Y^k (0)$. 
Averaging $D^k$ across all students, the average treatment effect (ATE) defined by $ATE=\mathbb{E}[Y(1)-Y(0)]$ is obtained. 
Similarly, averaging $D^k$ over the students whose input variable ${\bf X}$ equals ${\bf x}$,  
CATE is defined by 
$\tau({\bf x})=\mathbb{E}[Y(1)-Y(0)|{\bf X}={\bf x}]$ \cite{rubin}. 
To estimate ATE or CATE, we used T-learner based on Causal Inference. 

The notation used in the analysis is defined as follows: 
\begin{itemize}
  \item $R_1$: the set of 91 students who used F2F in 1Q. 
  \item $R_0$: the set of 1,298 students who did not use F2F in 1Q. 
  \item $i$: $i$-th student in $R_1$.
  \item $j$: $j$-th student in $R_0$.
  \item $k$: $k$-th out of $N=$1,389 students who took the 1Q Diff. regular examination. 
  \item $Y^{k}(1)$: $k$-th student's deviation on the 1Q Diff. regular examination if student $k$ used F2F . 
  \item $Y^{k}(0)$: $k$-th student's deviation on the 1Q Diff. regular examination if student $k$ did not use F2F. 
  \item $({\bf X}^{i,obs}, Y^{i,obs}(1))$ ($i \in R_1$): $i$-th student's observations comprising the input variable and deviation on the 1Q Diff. regular examination.
  \item $({\bf X}^{j,obs}, Y^{j,obs}(0))$ ($j \in R_0$): $j$-th student's observations comprising the input variable and deviation on the 1Q Diff. regular examination.
  \item ${\bf X}^{k,obs}$ ($k=1,\cdots, N$): $k$-th student's input observation.
\end{itemize}


We estimated the treatment response function 
from the users' observations $({\bf X}^{i,obs}, Y^{i,obs}(1))$ ($i \in R_1$), 
$\mu_1({\bf x}) = \mathbb{E}[Y(1) | {\bf X} ={\bf x} ] $ using supervised learning or regression. 
We denote the estimated function as $\hat{\mu}_1({\bf x})$. 
Similarly, from the nonusers' observations $({\bf X}^{j,obs}, Y^{j,obs}(0))$ ($j \in R_0$), 
we estimated the control response function 
$\mu_0({\bf x}) = \mathbb{E}[Y(0) |{\bf X} ={\bf x}]$ and denote the estimated function as $\hat{\mu}_0({\bf x})$.

\subsection{T-learner with ${\bf X}, {\bf x} \in \mathbb{R}^1$} 
\label{sec:xr1}

\begin{figure}[t]
  \centering
  \includegraphics[scale=0.3]{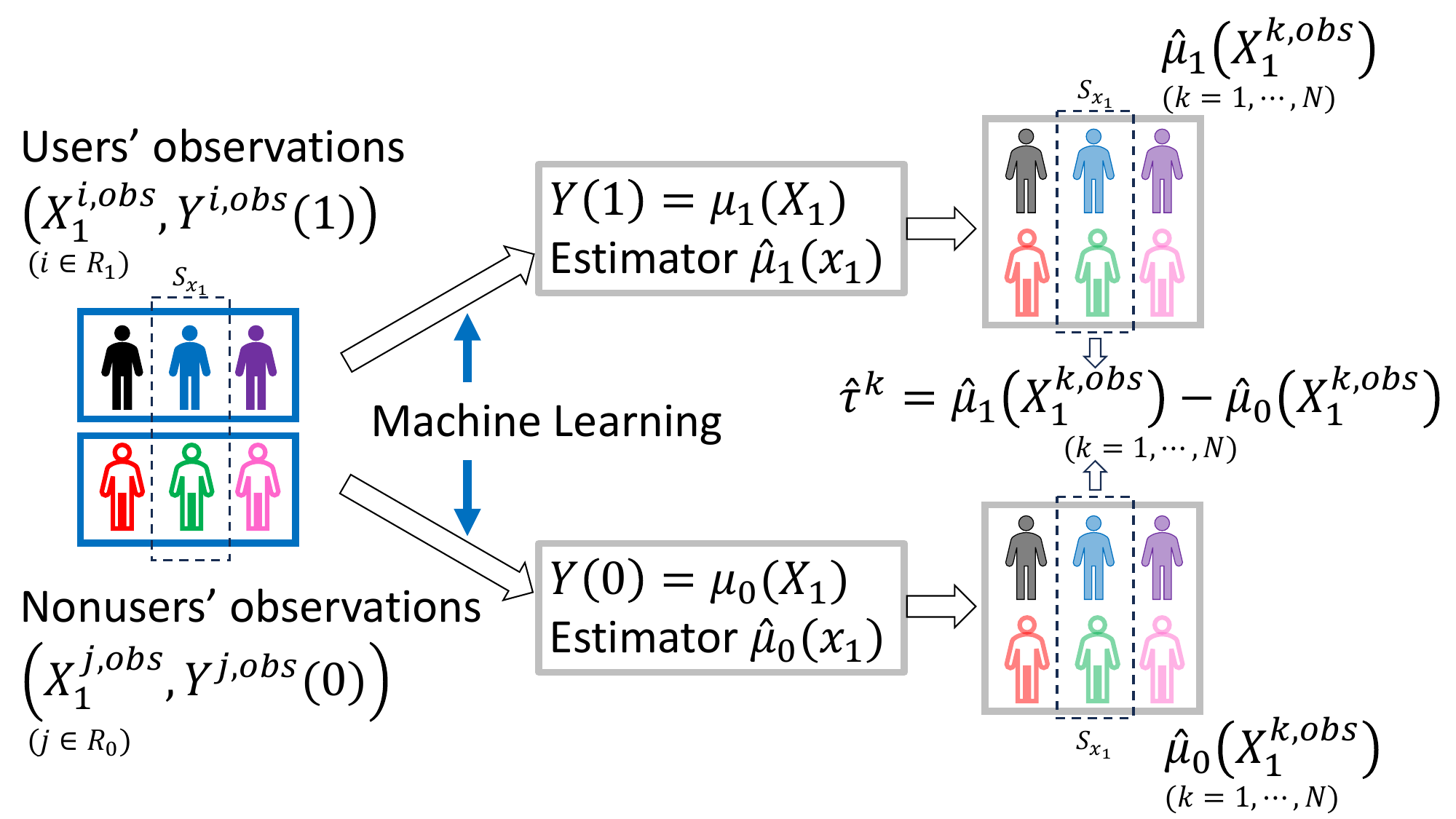}
  \caption{T-learner with $X_1$.}
  \label{fig:t1}
\end{figure}

Here, the input variable ${\bf X}$ is a one-dimensional vector ${\bf X}=(X_1)$. 
$X_1$ is the proficiency test deviation value, corresponding to a covariate. 
Fig. \ref{fig:t1} shows the framework of T-learner in the one-variable case of $X_1$.
The function $\hat{\mu}_1({\bf x})$ is estimated using the users' observations $(X_1^{i,obs}, Y^{i,obs}(1)) (i\in R_1)$. 
Similarly, the function $\hat{\mu}_0({\bf x})$ is estimated using the nonusers' observations $(X_1^{j,obs}, Y^{j,obs}(0)) (j\in R_0)$.
Among students who took the 1Q Diff. regular examination, let $S_{x_1}$ be the set of students whose proficiency test deviation value is $x_1$. 
The total number of students is $N =\sum_{x_1} |S_{x_1}| =$ 1,389. 
Here, $|S_{x_1}|$ represents the number of students included in $S_{x_1}$. 
Let $X_1^{k,obs}$ be the observed proficiency test deviation value for student $k$. 
Substituting $X_1^{k,obs}$ into these estimated functions, 
$\hat{\mu}_1(X_1^{k,obs})$ is obtained as the estimated deviation value of 1Q Diff. if student $k$ uses F2F;
$\hat{\mu}_0(X_1^{k,obs})$ is the estimated deviation value of 
1Q Diff. if student $k$ does not use F2F. 
The individual treatment effect for student $k$ is estimated as 
$\hat{\tau}^k = \hat{\mu}_1(X_1^{k,obs}) - \hat{\mu}_0 (X_1^{k,obs}) (k = 1, \cdots, N)$, 
from which CATE estimator $\hat{\tau} (x_1)$ for a proficiency test deviation value of $x_1$ is given as follows:
\begin{equation}
\hat{\tau} (x_1 )=\hat{\mu}_1 (x_1 )-\hat{\mu}_0 (x_1 )=\hat{\mu}_1 (X_1^{k,obs})-\hat{\mu}_0 (X_1^{k,obs})  (k\in S_{x_1}).    
\label{equ:tau}
\end{equation}
The CATE estimator $\hat{\tau}(x_1)$ was originally averaged over students belonging to $S_{x_1}$ \cite{rubin, kunzel}. 
However, the estimated functions are one-variable functions of $x_1$. 
The values do not change before and after averaging over the students in $S_{x_1}$. 
Therefore, we use this notation before determining the average.
Next, by averaging $\hat{\tau}^k$ over all students, ATE is calculated as follows: 
\begin{equation}
ATE=\frac{1}{N} \sum _{k=1}^N \{\hat{\mu}_1 (X_1^{k,obs})-\hat{\mu}_0 (X_1^{k,obs}) \}. 
\label{equ:ATE}
\end{equation}

Average Treatment Effect on the Treated (ATT) for F2F users 
and Average Treatment Effect on the Untreated (ATU) for nonusers 
are calculated as follows\cite{rubin, kunzel}:
\begin{eqnarray}
\label{equ:ATT}
ATT&=&\frac{1}{|R_1 |} \sum _{i\in R_1} \{Y^{i,obs} (1)-\hat{\mu}_0 (X_1^{i,obs}) \}, \\
ATU&=&\frac{1}{|R_0 |} \sum _{j\in R_0} \{\hat{\mu}_1 (X_1^{j,obs})-Y^{j,obs} (0) \}, 
\label{equ:ATU}
\end{eqnarray}
where $|R_1 |$= 91 and $|R_0 |$= 1,298.

In this paper, $\hat{\mu}_1(x_1)$ and $\hat{\mu}_0(x_1)$ are learned by means of random forests (regression) \cite{forest}. 
The depth of the trees was set to 2 for both 
$\hat{\mu}_1(x_1)$ and $\hat{\mu}_0(x_1)$, 
since the scores are
highest at that depth. 
We implemented these processes using
Python's scikit-learn.

\subsection{T-learner with ${\bf X}, {\bf x} \in \mathbb{R}^2$} 
\label{sec:xr2}

\begin{figure}[t]
  \centering
 \includegraphics[scale=0.3]{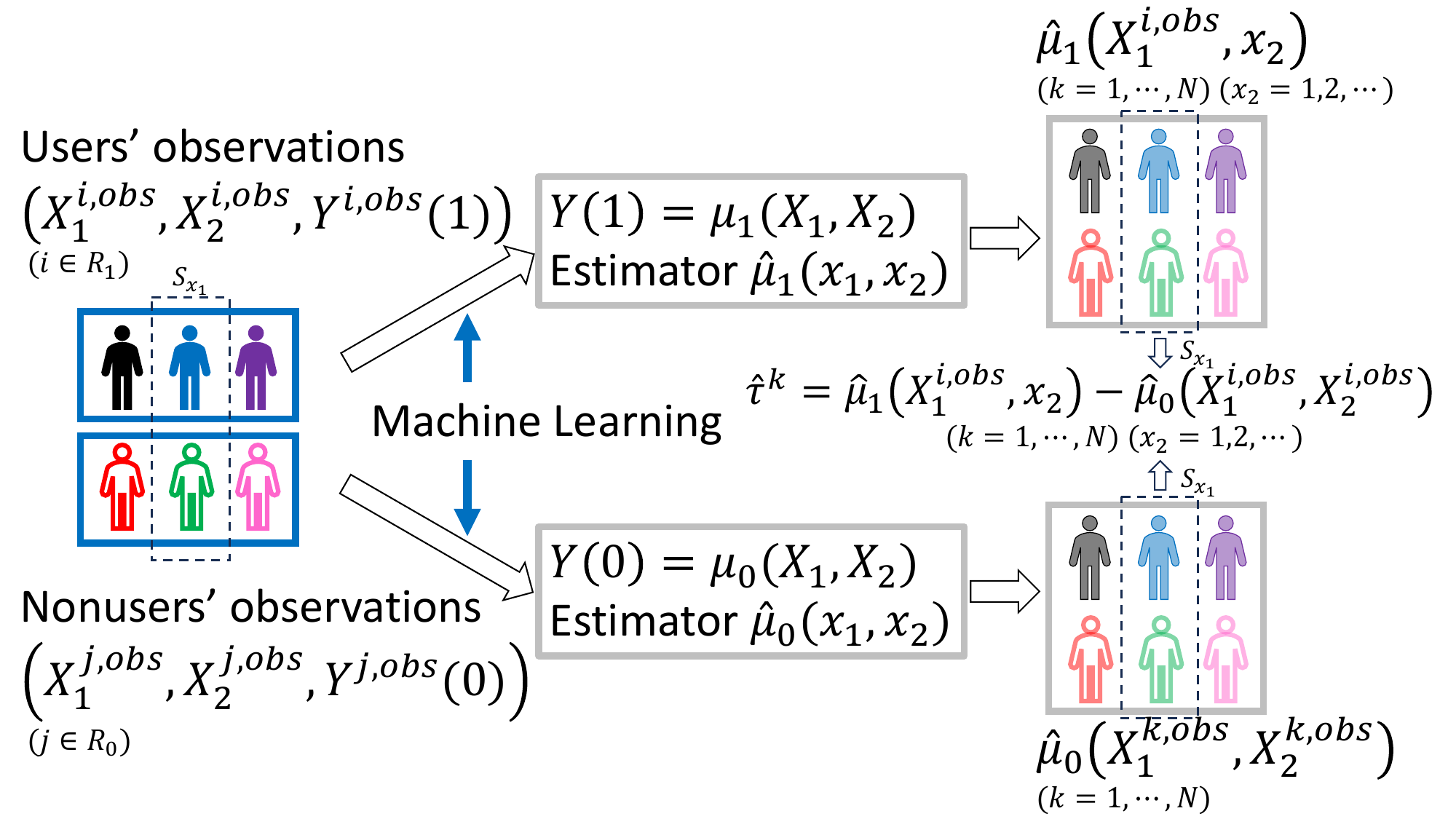}
  \caption{T-learner with $X_1$ and $X_2$.}
  \label{fig:t2}
\end{figure}

Extending the framework of T-learner, 
we now consider the case where the input variable ${\bf X}$ is a 2-dimensional vector ${\bf X}=(X_1, X_2)$.
The variable $X_2$ is the number of F2F sessions, corresponding to the number of treatments. 
Let $X_2^{k,obs}$ be the observed number of F2F sessions for student $k$. 

Fig. \ref{fig:t2} shows the framework of T-learner in the two-variable case of $X_1$ and $X_2$.
The function $\hat{\mu}_1({\bf x})$ is estimated using the users' observations $(X_1^{i,obs}, X_2^{i,obs}, Y^{i,obs}(1))$ $(i\in R_1)$. 
The function $\hat{\mu}_0({\bf x})$ is estimated using the nonusers' observations $(X_1^{j,obs}, X_2^{j,obs}, Y^{j,obs}(0))$ $(j\in R_0)$. 
The object is to optimize $\hat{\mu}_0 ({\bf x})$ and $\hat{\mu}_1 ({\bf x})$ with $X_1$ and $X_2$. 
The reason for considering two variables is as follows: 
multiple regression analysis indicates the potential effect of the number of F2F (Appendix \ref{sec:dis}). 
Moreover, the split by {\em Proficiency} in the decision tree (Appendix \ref{sec:tree}) is at the root of the tree, indicating that {\em Proficiency} has a stronger impact than {\em F2F}.
Consequently, we optimize $\hat{\mu}_0 ({\bf x})$ and $\hat{\mu}_1 ({\bf x})$ with $X_1$ and $X_2$. 
Estimation of these functions was optimized using random forests (regression) with depth 2 as in Appendix \ref{sec:xr1}, implemented with Python's scikit-learn. 

Multiple regression analysis indicates the potential effect of the number of F2F sessions (Appendix \ref{sec:dis}). Therefore, we explicitly add the number of F2F $X_2$ as an input variable. 
The input variable ${\bf X}$ is a 2-dimensional vector ${\bf X}=(X_1,X_2)$.



Similar to the previous case in Appendix \ref{sec:xr1}, $\hat{\mu}_0(X_1^{k,obs}, Y^{k,obs}(1))$ is the 1Q Diff. deviation value if student $k$ does not use F2F. 
The estimated function $\hat{\mu}_0(x_1, x_2)$ is learned from the nonusers' observations, 
where $X_2^{j,obs}=0 (j\in R_0)$. 
Thus, $\hat{\mu}_0(x_1, x_2)$ is expected to be independent of $x_2$. 
In fact, we confirmed that for any student $k$ and any $X_2$, $\hat{\mu}_0(X_1^{k,obs}, X_2^{k,obs})=\hat{\mu}_0(X_1^{k,obs}, x_2)=\hat{\mu}_0(X_1^{k,obs}, 0)$ holds. 
This means that $\hat{\mu}_0(x_1, x_2)$ is valid 
and equivalent to a base treatment response function in multi-level treatment meta-learners\cite{acha}. 

Next, the function  
$\hat{\mu}_1(x_1, x_2)$ is learned from the users' observations, where 
the users' number of sessions of F2F $X_2^{i,obs}\ge 1 (i\in R_1)$. 
Thus, $\hat{\mu}_1(x_1, x_2)$ is defined in  the domain $x_2 \ge 1$. 
Consequently, we cannot substitute nonusers' $X_2^{j,obs}=0$ into $\hat{\mu}_1(x_1, x_2)$. 
Rather, instead of estimating 91 $\hat{\mu}_1(X_1^{i, obs}, X_2^{i, obs}) (i \in R_1)$ by substituting 91 users' observations, 
we calculate 1,389 estimated $\hat{\mu}_1(X_1^{k, obs}, x_2) (k=1, \cdots, N)$ by substituting the variable $x_2$ in Eq. \eqref{equ:phi} artificially. 
For example, in the case of only one F2F session, $x_1=1$ is substituted. 
Thus, 1,389 predictive $\hat{\mu}_1(X_1^{k, obs}, 1) (k=1, \cdots, N)$ values are obtained. 
Together with the relation $\hat{\mu}_0(X_1^{k,obs}, X_2^{k,obs})=\hat{\mu}_0(X_1^{k,obs}, 0)$, 
the difference $\hat{\mu}_1(X_1^{k,obs}, 1) -\hat{\mu}_0(X_1^{k,obs}, X_2^{k,obs}) = \hat{\mu}_1(X_1^{k,obs}, 1) -\hat{\mu}_0(X_1^{k,obs}, 0) (k=1, \cdots, N) $ indicates that there are 1,389 predictive individual treatment effects if one F2F session is used.
Similarly, we can obtain 
the differences $\hat{\mu}_1(X_1^{k, obs}, 2)-\hat{\mu}_0(X_1^{k,obs}, X_2^{k,obs})$, $\hat{\mu}_1(X_1^{k, obs}, 3)-\hat{\mu}_0(X_1^{k,obs}, X_2^{k,obs})$, 
$\cdots$, and 
$\hat{\mu}_1(X_1^{k, obs}, x_2)-$ \\
$\hat{\mu}_0(X_1^{k,obs}, X_2^{k,obs})$ 
$ (k=1, \cdots, N)$. 
The difference $\hat{\mu}_1(X_1^{k,obs}, x_2) -\hat{\mu}_0(X_1^{k,obs}, X_2^{k,obs}) = \hat{\mu}_1(X_1^{k,obs}, x_2) -\hat{\mu}_0(X_1^{k,obs}, 0) (k=1, \cdots, N) $ indicates the predictive individual treatment effects if each student $k$ participates in $x_2$ F2F sessions. 
Averaging this individual treatment effect over all the students in $S_{x_1}$, we propose the 
CATE estimator $\varphi(x_1, x_2)$ of Eq. \eqref{equ:phi} in Section \ref{sec:resultxr1}. 
Furthermore, for the same reason given for Eq. \eqref{equ:tau}, 
$\varphi (x_1,x_2 )$, left-hand side of Eq. \eqref{equ:phi} in Section \ref{sec:resultxr2},  
is equal to a summand in Eq. \eqref{equ:phi}, that is

\begin{eqnarray}
\varphi (x_1,x_2 )&=&\hat{\mu}_1 (X_1^{k,obs},x_2 )-\hat{\mu}_0 (X_1^{k,obs}, X_2^{k,obs})   \quad (x_2=1,2,\cdots ) \quad (k\in S_{x_1 }).
\label{equ:phi2}
\end{eqnarray}


Using the observations of users $Y^{i,obs}(1)$, we calculate the average treatment effect for users in the two-variable case $ATT_2$ as follows: 
\begin{equation}
ATT_2=\frac{1}{|R_1 |} \sum_{i\in R_1} \{ {Y^{i,obs} (1)-\hat{\mu}_0 (X_1^{i,obs}, X_2^{i,obs})} \} (i\in R_1). 
\end{equation}
Based on the relationship $\hat{\mu}_0 (X_1^{k,obs}, X_2^{ k,obs}) =\hat{\mu}_0 (X_1^{k,obs}, 0)$, $ATT_2$ is expected to be equal to the value of $ATT$ in Eq. (\ref{equ:ATT}). 

\section{Discussion of the Effect of the number of F2F sessions }
\label{sec:dis}
\begin{figure}[h]
  \centering
  \includegraphics[scale=0.6]{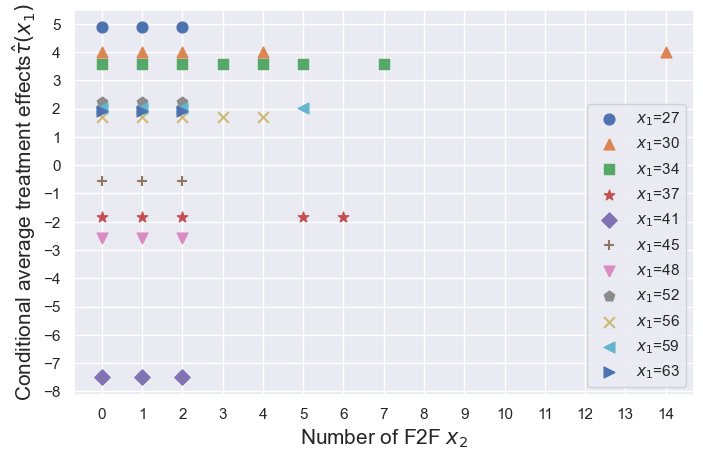}
  \caption{CATE estimator $\hat{\tau}(x_1)$ with the number of F2F sessions.}
  \label{fig:catef2f}
\end{figure}

Fig. \ref{fig:catef2f} shows a scatter plot of the data 
$(X_1^{k,obs},X_2^{k,obs},\hat{\tau} (X_1^{k,obs})) (k=1,\cdots ,N)$, 
which is a combination of 
$(X_1^{k,obs},\hat{\tau} (X_1^{k,obs}))$ (plotted as $\textcolor{red}{\bullet}$ in Fig. \ref{fig:CATE}) and the actual observed number of F2F sessions $X_2^{k,obs}$. 
$X_2^{k,obs}$ is plotted on the horizontal axis; $\hat{\tau} (X_1^{k,obs})$ is plotted on the vertical axis, with marker types corresponding to different groups $S_{x_1}$ of $X_1^{k,obs}=x_1$. 
For the same $x_1$ (same color and same shape markers in Fig. \ref{fig:catef2f}), 
the values of $\hat{\tau} (x_1 )$ are the same (vertical axis); 
however, some students may have different values of $X_2^{k,obs}$ (horizontal axis). 
Regarding the values of the CATE estimator $\hat{\tau} (x_1)$, 
data with values below $-2$ exist for 0--2 F2F sessions, 
but such data disappear for 3 sessions or more.
For 3 or 4 sessions, 7 sessions, and 14 sessions, all the data points have positive values. 
This suggests that more F2F sessions tend to result in a larger $\hat{\tau} (x_1)$. 
Thus, there is the possibility of an effect depending on the number of F2F sessions. 

In the machine learning approaches discussed in Appendix \ref{sec:xr1} and Section \ref{sec:resultxr1}, 
$\hat{\tau} (x_1)$ in Eq. \eqref{equ:tau} is considered as one-variable function of $x_1$.
The number of F2F sessions is not explicitly included.
However, to explore the potential effect of the number of F2F sessions as seen in Fig. \ref{fig:catef2f}, 
we use multiple regression analysis by explicitly including the data on the number of sessions. 
Using the dataset $(X_1^{k,obs},X_2^{k,obs}, \hat{\tau} (X_1 ^{k,obs})) (k=1,\cdots ,N)$, 
we obtain the regression equation $y=0.10x_1+0.44x_2-4.8$.  
Here, $\hat{\tau} (x_1)$ is equivalent to the output $y$. 
$x_1$ and $X_2=x_2$ are input variables.
The partial regression coefficient of $x_2$ is positive. 
This means that $\hat{\tau} (x_1 )$ becomes larger as the number of F2F sessions increases, 
implying the potential impact of the number of F2F sessions. 

\vspace{3cm}

\section{3D surface plot of $\varphi(x_1, x_2)$}
\label{sec:3D}

\begin{figure}[h]
  \centering
  \hspace{0cm}
 \includegraphics[scale=0.95]{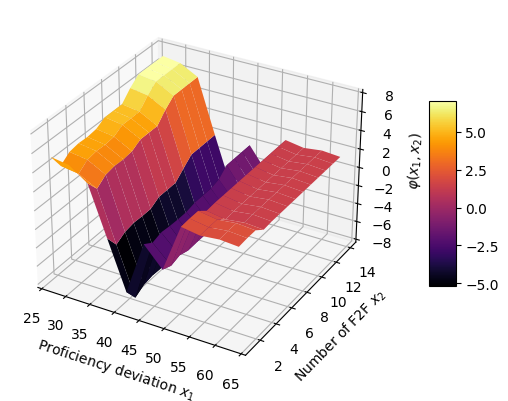}
  \caption{3D surface plot of $\varphi(x_1, x_2)$.}
\label{fig:3D}  
\end{figure}


\begin{thebibliography}{00}
\bibitem{Ala}Ahmed M. Alaa and Mihaela van der Schaar. 2017. Bayesian inference of individualized treatment effects using multi-task gaussian processes. In {\it Proceedings of the 31st International Conference on Neural Information Processing Systems} NIPS'17 (2017), 3427-3435. 

\bibitem{acha} Naoufal Acharki, Ramiro Lugo, Antoine Bertoncello, and Josselin Garnier. 2023. Comparison of meta-learners for estimating multi-valued treatment heterogeneous effects. In {\it Proceedings of the 40th International Conference on Machine Learning} PMLR 202 (2023), 91-132.

\bibitem{ba}Berardino Barile, Marco Forti, Alessia Marrocco, and Angelo Castaldo. 2024. Causal impact evaluation of occupational safety policies on firms' default using machine learning uplift modelling. {\it Scientific Reports} 14 (2024), article number: 10380. 

\bibitem{Breiman}Leo Breiman, Jerome H. Friedman, Richard A. Olshen, and Charles J. Stone. 1984. {\it Classification and Regression Trees}. Wadsworth International Group.

\bibitem{forest}Leo Breiman. 2001. Random Forests. {\it Machine Leaning} 45, 1 (2001), 5-32.

\bibitem{caronb} Alberto Caron, Gianluca Baio, and Ioanna Manolopoulou. 2020. Estimating individual treatment effects using non-parametric regression models: a review. {\it Journal of Royal Statistical Society Series A} 185, 3 (2022), 1115-1149.

\bibitem{fujihara}Sho Fujihara. 2019, Causal inference in studies of sociology of education. {\it Sociological Theory and Methods} 34, 1 (2019), 65-77. (in Japanese)

\bibitem{harada}Shonosuke Harada and Hisashi Kashima. 2021. Graphite: estimating individual effects of graph-structured treatments. In {\it Proceedings of the 30th ACM International Conference on Information and Knowledge Management}. 659-668. 

\bibitem{hu}Liangyuan Hu, Chenyang Gu, Michael Lopez, Jiayi Ji, and Juan Wisnivesky. 2020. Estimation of causal effects of multiple treatments in observational studies with a binary outcome. {\it Statistical methods in medical research} 29, 11 (2020),3218-3234.

\bibitem{imbens}Guido W. Imbens. 2000. The role of propensity score in estimating dose-response function. {\it Biometrika} 87, 3 (2000), 706-710. 

\bibitem{rubin}Guido W. Imbens and Donald B. Rubin. 2015. {\it Causal inference in statistics, social, and biomedical sciences}. Cambridge University Press.

\bibitem{kell}Bryan Keller, Zach Branson. 2024. Defining, identifying, and estimating causal effects with the potential outcomes framework: A review for education research. {\it Asia Pacific Education Review} May 2024.  https://doi.org/10.1007/s12564-024-09957-2.

\bibitem{knu}Michael C. Knaus. 2022. Double machine learning-based programme evaluation under unconfoundedness. {\it The Econometrics Journal} 25, 3 (2022), 602-627.

\bibitem{knu2}Michael C. Knaus,  Michael Lechner, and Anthony Strittmatter. 2022. A Heterogeneous employment effect of job search programmes: A machine learning approach. {Journal of Human Resources} 57, 2 (2022), 597-636.

\bibitem{kunzel} S\"{o}ren R. K\"{u}nzel, Jasjeet S. Sekhon, Peter J. Bickel, and Bin Yu. Metalearners for estimating heterogeneous treatment effects using machine learning. In {\it Proceedings of the national academy of sciences} 116, 10 (2019), 4156-4165.

\bibitem{lech}Michael Lechner. 2001. Identification and estimation of causal effects of multiple treatments under the conditional independence assumption. In {\it M. Lechner and F. Pfeiffer (eds.) Econometric Evaluation of Labour Market Policies, Heidelberg: Physica}. 43-58.

\bibitem{lin}Lin Lin, Yeying Zhu, and Liang Chen. 2019. Causal inference for multi-level treatments with machine-learned propensity scores. {\it Health Services and Outcomes Research Methodology} 19, 2 (2019), 106-126.

\bibitem{luo}Huishi Luo, Fuzhen Zhuang, Ruobing Xie, Hengshu Zhu, Deqing Wang, Zhulin An, and Yongjun Xu. 2024. A survey on causal inference for recommendation. {\it The Innovation} 5, 2 (2024), 100590. https://doi.org/10.1016/j.xinn.2024.100590


\bibitem{tnagai1}Tomoko Nagai, Motoki Kino, Tetsuo Hosoya, Hirohisa Takahashi, Takanao Tsuyuki, Takayuki Muto. 2022. Analysis of educational effects at academic support center in Kogakuin university based on decision tree method. {\it Physics Education in University} 28, 2 (2022), 94-97.(in Japanese)
\bibitem{tnagai2} Tomoko Nagai, Kensaku Kinjo, Kengo Kawamura, Tomoya Nakamura, Takayuki Okuda, Yuichiro Sato, Jun-ichi Mukuno, Shin Kikuta, and Naoto Kumano-go. 2023. Educational effects by the division of mathematics of Kogakuin university academic support center: decision tree analysis. {\it Japanese Society for Engineering Education} 71, 3 (2023), 112-116.(in Japanese)
\bibitem{tnagai3}Tomoko Nagai, Takayuki Okuda, Tomoya Nakamura, Yuichiro Sato, Yusuke Sato, Kensaku Kinjo, Kengo Kawamura, Shin Kikuta, and Naoto Kumano-go. 2024. Evaluation of Educational effects in mathematics by means of T-learner based upon causal inference. {\it Japanese Society for Engineering Education} in press. (in Japanese)

\bibitem{ney} Jerzy Splawa-Neyman, D. M. Dabrowska, T. P. Speed. 1990. On the application of probability theory to agricultural experiments. Essay on principles. Section 9. {\it Statistical Science} 5 (1990),465-472. 

\bibitem{yutas1} Yuta Saito, Hayato Sakata, and Kazuhide Nakata. 2020. Cost-Effective and Policy Optimaization Algorithm for Uplift Modeling with Multiple Treatments. In {\it Proceedings of the 2020 SIAM International Conference on Data Mining}. 406-414.

\bibitem{yutas3}Yuta Saito. 2024. {\it CounterFactual Machine Learning}. Gijutuhyoronsya. (in Japanese) 



\bibitem{rubin2} Donald B. Rubin. 1974. Estimating causal effects if treatment in randomized and nonrandomized studies. {\it Journal of Educational Psychology} 66, 5 (1974), 688-701.





\end{thebibliography}
\end{document}